\documentclass[pdflatex,sn-nature]{sn-jnl}

\usepackage{geometry}
\usepackage{graphicx}%
\usepackage{multirow}%
\usepackage{amsmath,amssymb,amsfonts}%
\usepackage{amsthm}%
\usepackage{mathrsfs}%
\usepackage[title]{appendix}%
\usepackage{xcolor}%
\usepackage{textcomp}%
\usepackage{manyfoot}%
\usepackage{booktabs}%
\usepackage{algorithm}%
\usepackage{algorithmicx}%
\usepackage{algpseudocode}%
\usepackage{listings}%

\usepackage{hyperref}
\usepackage{bm}
\usepackage{comment}

\unnumbered

\begin{document}

\title{Stacking-induced ferroelectricity in tetralayer graphene}


\author*[1,2]{\fnm{Amit} \sur{Singh}}\email{amitletit@gmail.com}

\author[3,4]{\fnm{Shuigang} \sur{Xu}}

\author[1,2]{\fnm{Patrick Johansen} \sur{Sarsfield}}

\author[1,2]{\fnm{Pablo Díaz} \sur{Núñez}}

\author[1,2]{\fnm{Ziwei} \sur{Wang}}

\author[1,2]{\fnm{Sergey} \sur{Slizovskiy}}

\author[1,2]{\fnm{Nicholas} \sur{Kay}}

\author[5]{\fnm{Jun} \sur{Yin}}

\author[1,2]{\fnm{Yashar} \sur{Mayamei}}

\author[6]{\fnm{Takashi} \sur{Taniguchi}}

\author[6]{\fnm{Kenji} \sur{Watanabe}}

\author[1,2]{\fnm{Qian} \sur{Yang}}

\author[7]{\fnm{Kostya S.} \sur{Novoselov}}

\author[1,2]{\fnm{Vladimir I.} \sur{Fal'ko}}

\author*[1,2]{\fnm{Artem} \sur{Mishchenko}}\email{artem.mishchenko@manchester.ac.uk}

\affil[1]{\orgdiv{Department of Physics and Astronomy}, \orgname{University of Manchester}, \orgaddress{\street{Oxford Road}, \city{Manchester}, \postcode{M13 9PL}, \country{UK}}}

\affil[2]{\orgdiv{National Graphene Institute}, \orgname{University of Manchester}, \orgaddress{\street{Oxford Road}, \city{Manchester}, \postcode{M13 9PL}, \state{State}, \country{UK}}}

\affil[3]{\orgdiv{Key Laboratory for Quantum Materials of Zhejiang Province, Department of Physics, School of Science}, \orgname{Westlake University}, \orgaddress{\street{18 Shilongshan Road}, \city{Hangzhou}, \postcode{310024}, \state{Zhejiang}, \country{China}}}

\affil[4]{\orgdiv{Institute of Natural Sciences}, \orgname{Westlake Institute for Advanced Study}, \orgaddress{\street{18 Shilongshan Road}, \city{Hangzhou}, \postcode{310024}, \state{Zhejiang}, \country{China}}}

\affil[5]{\orgdiv{State Key Laboratory of Mechanics and Control for Aerospace Structures, Key Laboratory for Intelligent Nano Materials and Devices of Ministry of Education, Institute for Frontier Science}, \orgname{Nanjing University of Aeronautics and Astronautics}, \city{Nanjing}, \state{Jiangsu}, \country{China}}

\affil[6]{\orgname{National Institute for Materials}, \orgaddress{\street{Namiki 1-1, Tsukuba}, \city{Ibaraki}, \postcode{305-0044}, \country{Japan}}}

\affil[7]{\orgname{Institute for Functional Intelligent Materials, National University of Singapore}, \orgaddress{\street{Block S9, Level 9, 4 Science Drive 2}, \city{Singapore}, \postcode{117544}, \country{Singapore}}}


\abstract{Recent studies have reported emergent ferroelectric behavior in twisted or moir\'e-engineered graphene-based van der Waals heterostructures \cite{zhang2024electronic, zheng2020unconventional, zheng2023electronic, niu2022giant, yan2023moire, ren2023ferroelectric,waters2024origin, chen2024anomalous, klein2023electrical, wang2022ferroelectricity, chen2024selective,niu2025ferroelectricity, lin2025room}, yet the microscopic origin of this effect remains under debate. Pristine mono- or few-layer graphene lacks a permanent dipole due to its centrosymmetric lattice, making the emergence of ferroelectricity unlikely. However, mixed-stacked graphene, such as the ABCB tetralayer configuration, breaks both inversion and mirror symmetry and has been theoretically predicted to support electrically switchable dipoles \cite{fischer2024spin, sarsfield2024substratetemperaturemagneticfield, garcia2023mixed}. ABCB graphene represents the simplest natural graphene polytype exhibiting intrinsic out-of-plane polarization, arising from asymmetric charge carrier distribution across its layers \cite{atri2024spontaneous, zhou2024inversion, wirth2022experimental}. Here, we report robust ferroelectric behavior in dual-gated, non-aligned ABCB tetralayer graphene encapsulated in hexagonal boron nitride. The device exhibits pronounced hysteresis in resistance under both top and bottom gate modulation, with the effect persisting up to room temperature. This hysteresis originates from reversible layer-polarized charge reordering, driven by gate-induced transitions between ABCB and BCBA stacking configurations -- without requiring moir\'e superlattices. Our findings establish stacking-order-induced symmetry breaking as a fundamental route to electronic ferroelectricity in graphene and open pathways for non-volatile memory applications based on naturally occurring mixed-stacked multilayer graphene.}




\maketitle

\section{Main}\label{sec1}
The discovery of ferroelectricity in two-dimensional (2D) van der Waals (vdW) materials has opened new avenues for studying emergent quantum phases and developing next-generation electronic and memory technologies \cite{wang2023towards}. Conventional ferroelectricity typically arises from the displacement of ionic sublattices in non-centrosymmetric crystals. In contrast, recent studies have uncovered unconventional ferroelectricity in moiré superlattices formed by graphene or transition metal dichalcogenides (TMDs), where spontaneous electric polarization originates from purely electronic mechanisms rather than lattice distortions \cite{zhang2024electronic,zheng2020unconventional,zheng2023electronic}. This so-called \textit{electronic ferroelectricity} enables reversible polarization switching and non-volatile memory states, with high tunability and minimal energy dissipation -- key features for ultra-low-power logic and neuromorphic applications \cite{niu2022giant,yan2023moire,waters2024origin}. Achieving stable and reproducible control over these effects remains an active challenge, particularly in systems where ferroelectricity is emergent rather than structurally encoded.

Graphene, owing to its centrosymmetric lattice and absence of a permanent dipole, was long considered an unlikely platform for ferroelectricity. However, moiré engineering has enabled the realization of ferroelectric-like behavior in systems like Bernal-stacked bilayer graphene (BLG) encapsulated by hexagonal boron nitride (hBN), or in magic-angle twisted bilayer graphene \cite{chen2024anomalous, zheng2023electronic, niu2022giant, klein2023electrical}. In these systems, the interfacial moiré potential modifies the low-energy electronic bands, leading to asymmetric interlayer charge transfer and spontaneous charge polarization. This results in characteristic hysteresis in resistance as a function of gate voltage, a hallmark of electronic ferroelectricity, which is driven by electronic reconstruction rather than ionic displacement. While such moiré-induced effects are now well-established in twisted bilayers and heterostructures, emerging evidence suggests that similar hysteresis can also arise in multilayer graphene without moiré patterns. In particular, the role of stacking order as an intrinsic symmetry-breaking mechanism -- capable of inducing ferroelectricity even in the absence of twist or alignment -- remains an underexplored but promising direction \cite{atri2024spontaneous, zhou2024inversion, wirth2022experimental}.

Ferroelectricity in two-dimensional (2D) graphene-based field effect transistors (FETs) is highly attractive for technological applications due to the inherently weak electrostatic screening and excellent gate tunability of charge carriers in these systems \cite{yan2023moire, beck2020spiking}. Such properties make graphene-based FETs promising candidates for non-volatile memory, where the ferroelectric effect can be harnessed to implement on/off states via gate-controlled hysteresis around the charge neutrality point (CNP).

Intrinsic Bernal stacked multilayer graphene does not exhibit ferroelectricity. In contrast, rhombohedral-stacked multilayers, such as trilayer \cite{winterer2024ferroelectric} and pentalayer \cite{han2023orbital} graphene, have been reported to show a weak ferroelectric effect at very low temperatures - attributed to strong electronic correlations in these systems. More pronounced effects have recently been observed in a variety of engineered systems, including: mono-, bi- , and trilayer graphene aligned with hBN forming layer-asymmetric moiré superlattices \cite{zhang2024electronic, lin2025room}; bilayer graphene under displacement fields or alignment-induced symmetry breaking \cite{zheng2020unconventional, zheng2023electronic, niu2022giant, yan2023moire, ren2023ferroelectric}; multilayer moiré graphene \cite{waters2024origin}; twisted double bilayer graphene \cite{chen2024anomalous}; 
magic angle twisted bilayer graphene (MATBG) \cite{klein2023electrical, chen2024selective}; hBN intercalated graphene \cite{wang2022ferroelectricity}; and graphene/WSe$_2$ heterostructures \cite{niu2025ferroelectricity, waters2024origin}. Ferroelectric switching has also been demonstrated in devices where graphene is electrostatically coupled to semiconducting TMDs 
 \cite{sterbentz2024gating}.

In nearly all previously reported cases, ferroelectric-like behavior in graphene-based systems has been attributed to a bistable electronic ground state emerging from moiré-induced potentials. A characteristic feature of such systems is that hysteresis typically manifests under one gate only -- indicating that interfacial coupling or gate-induced layer asymmetry plays a more critical role than intrinsic stacking order alone.

\begin{figure}[ht]
  \centering
  \includegraphics[scale=0.7]{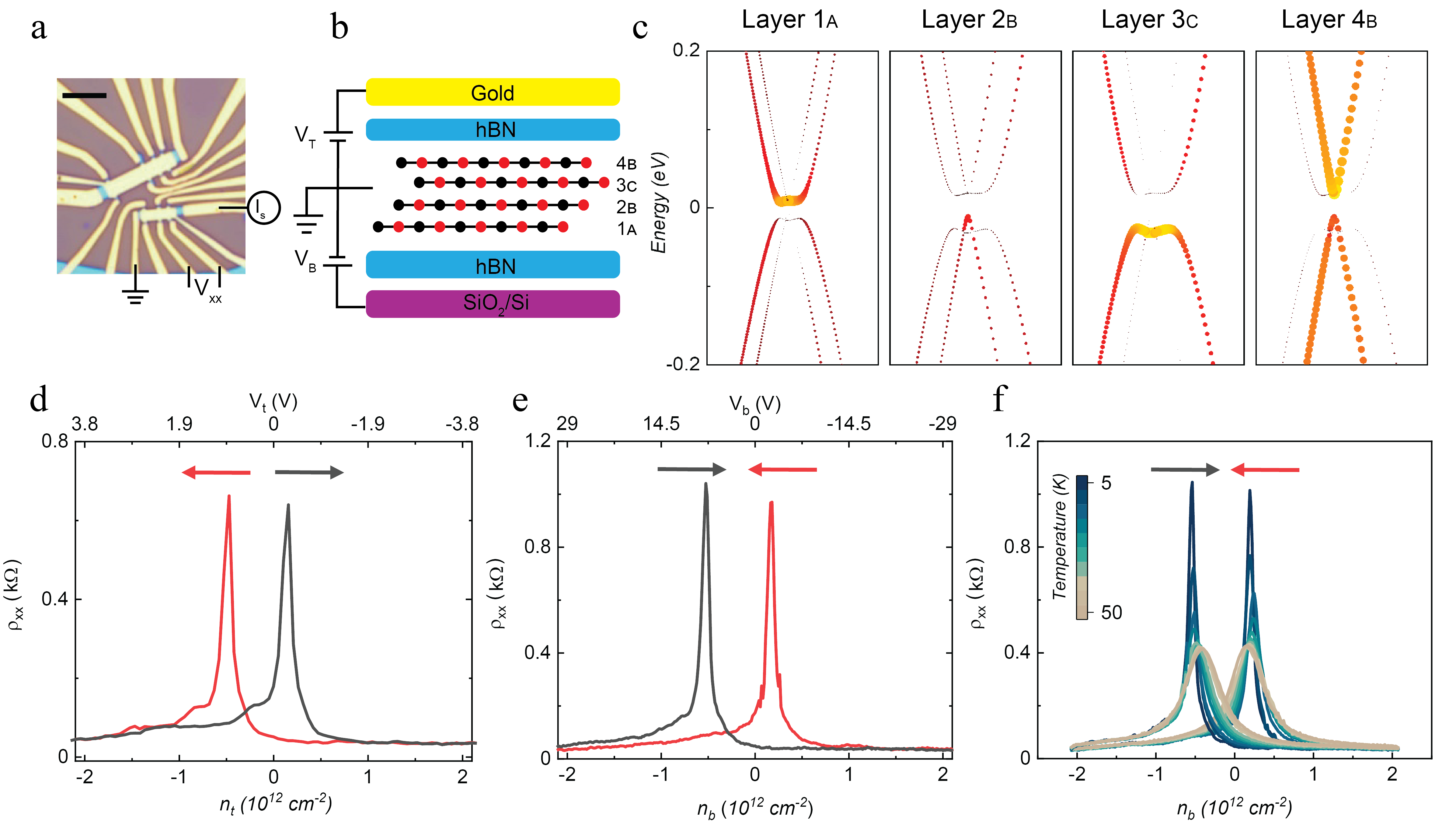}
  \caption{\textbf{Hysteresis in hBN-encapsulated tetralayer ABCB graphene device.} 
  \textbf{a,} Optical micrograph of the Hall bar device; annotations show the measurement setup, where $I_s$ is the excitation current source and $V_{xx}$ indicates the longitudinal voltage probes. The scale bar is $5\mu m$. 
  \textbf{b,}  Schematic of the hBN-encapsulated ABCB-stacked tetralayer graphene device. 
  \textbf{c,} Schematic representation of low energy band structure of ABCB graphene, illustrating electron localization in the $1_A$ layer, hole localization in the $3_C$ layer, and near charge neutrality in the $2_B$ and $4_B$ layers. Smaller black circles represent regions with low layer-projected density of states (LPDOS), while larger yellow circles indicate regions of high LPDOS. \textbf{d-e,}  Longitudinal resistivity $\rho_{xx}$ as a function of $n_t$ (d) and $n_b$ (e), measured at $n_b = 0$ and $n_t = 0$, respectively, at a temperature of 5 K. The top axes indicate the corresponding applied gate voltages. Black and red arrows denote forward and backward sweep directions, respectively. 
  \textbf{ f,} $\rho_{xx} (n_b)$ traces at temperatures from 5 to 50 K in 5 K steps.}
  \label{fig1}
\end{figure}

In this work, we study tetralayer graphene with an ABCB mixed stacking order, intentionally misaligned with both top and bottom hBN layers to eliminate moiré effects (device micrograph is shown in \textbf{Fig.\ref{fig1}a}, and the cross-section schematic in \textbf{Fig.\ref{fig1}b}). The ABCB configuration intrinsically breaks inversion and mirror symmetries \cite{zhou2024inversion}, and is predicted to host a small bandgap \cite{fischer2024spin}, correlated ground states even in the absence of a moiré superlattice \cite{fischer2024spin}, and  electrically switchable dipoles \cite{sarsfield2024substratetemperaturemagneticfield, garcia2023mixed, yang2023atypical}. These dipoles arise from asymmetric charge localization across the graphene layers -- specifically, electrons and holes residing in the \textbf{\textit{$1_A$}} and \textbf{\textit{$3_C$}} layers, respectively \textbf{(Fig.\ref{fig1}c)} -- leading to spontaneous out-of-plane polarization.

As the simplest naturally occurring graphene polytype exhibiting such broken symmetries, ABCB tetralayer graphene (ABCB) offers a minimal platform to probe spontaneous layer polarization and electronic symmetry breaking. Here, we report robust gate-controlled hysteresis in the longitudinal resistivity around the CNP, in the hBN-encapsulated ABCB graphene device.  The hysteresis persists up to room temperature, excluding thermally activated or low-energy trapping mechanisms, and occurs without requiring moiré superlattices or interlayer twist -- demonstrating that stacking order alone is sufficient to induce a ratchet-like ferroelectric response.

To isolate intrinsic effects, we fabricated dual-gated Hall bar device from hBN/ABCB/hBN heterostructure, with intentional angular misalignment between the graphene and both hBN interfacees \textbf{Fig.\ref{fig1} a,b}, see \textbf{Methods, ‘Device fabrication’} for details. hBN encapsulation ensures high carrier mobility \cite{yin2019dimensional, shi2020electronic}, while misalignment suppresses moiré reconstruction \cite{mullan2023mixing, yankowitz2012emergence, ponomarenko2013cloning}. Electrostatic gating via top and bottom gates enables independent tuning of carrier densities $n_t$, $n_b$, and control over the displacement field $D$, allowing systematic investigation of ferroelectric switching behavior under varied gating conditions.

\subsection{Hysteresis in ABCB graphene device}
We first measured the longitudinal resistivity, $\rho_{xx}$, as a function of $n_t$ at $n_b=0$ and $n_b$ at $n_t=0$ (\textbf{Fig.\ref{fig1}d,e}). Measurements were performed at $T=5$ K unless otherwise stated. In both cases, $\rho_{xx}$ peak at CNP shows large hysteresis between forward and backward sweep directions. We define forward (up) sweep direction as increasing charge carrier concentration $n$ (hole-to-electron doping), and backward (down) -- as decreasing $n$ (electron-to-hole doping).  The offset between the forward and backward sweeps is $\Delta n_t = 0.63 \cdot 10^{12}\text{cm}^{-2}$ for $n_t$ and $\Delta n_b = 0.7 \cdot 10^{12}\text{cm}^{-2}$ for $n_b$, and remains consistent across different sweep rates. 

In an ideal undoped device, the CNP would be located at $n_t = n_b = 0$. However, in our measurements, the peak position shifts: for the up sweep direction, $\rho_{xx}$ peaks in the electron-doped region for $n_t$ sweep, and in the hole-doped region for $n_b$ sweep, and vice versa for the down sweep direction. Additionally, the magnitude of $\rho_{xx}$ is larger when sweeping $n_b$ compared to $n_t$, and generally larger in the hole-doped regime than in the electron-doped regime. These asymmetries suggest that the out-of-plane electric displacement field differs between sweep directions, indicating layer-resolved charge localization. The hysteresis remains robust up to at least 50 K, with minimal change in offset. The main temperature dependence is a broadening of the $\rho_{xx}$ peak, consistent with thermal broadening at elevated temperatures (\textbf{Fig.\ref{fig1}f}). 

To understand the origin of the hysteresis and its dependence on gate polarity, we calculated the energy dispersion of the ABCB-stacked tetralayer graphene (see \textbf{Methods, ‘DFT calculations' and ‘SWMC Model for ABCB graphene system'} for details). The band structure along the high-symmetry path $\Gamma-M-K-\Gamma$ shows that four low-energy bands -- two conduction and two valence -- approach each other near the $K$-point (\textbf{Extended Data Fig.\ref{figSI_dispersion}}). Layer-resolved projections reveal that the conduction band states are primarily localized in layer $1_A$, while the valence band states are concentrated in layer $3_C$. The remaining Dirac-like bands reside in layers $2_B$ and $4_B$.

This spatial asymmetry explains the observed doping-dependent hysteresis. When sweeping $n_b$, 
the bottom gate preferentially modulates charge carriers in the nearby $1_A$ layer, requiring additional hole doping to compensate for localized excess electrons. Similarly, sweeping $n_t$ modulates the $3_C$ layer, leading to the opposite shift in the charge neutrality point. This layer-specific gating response underpins the observed reversal of peak positions between top and bottom gate sweeps (Extended Data Fig.\ref{figSI_dispersion}).

\subsection{Layer polarization in ABCB graphene}
\begin{figure}[ht]
  \centering
  \includegraphics[scale=0.8]{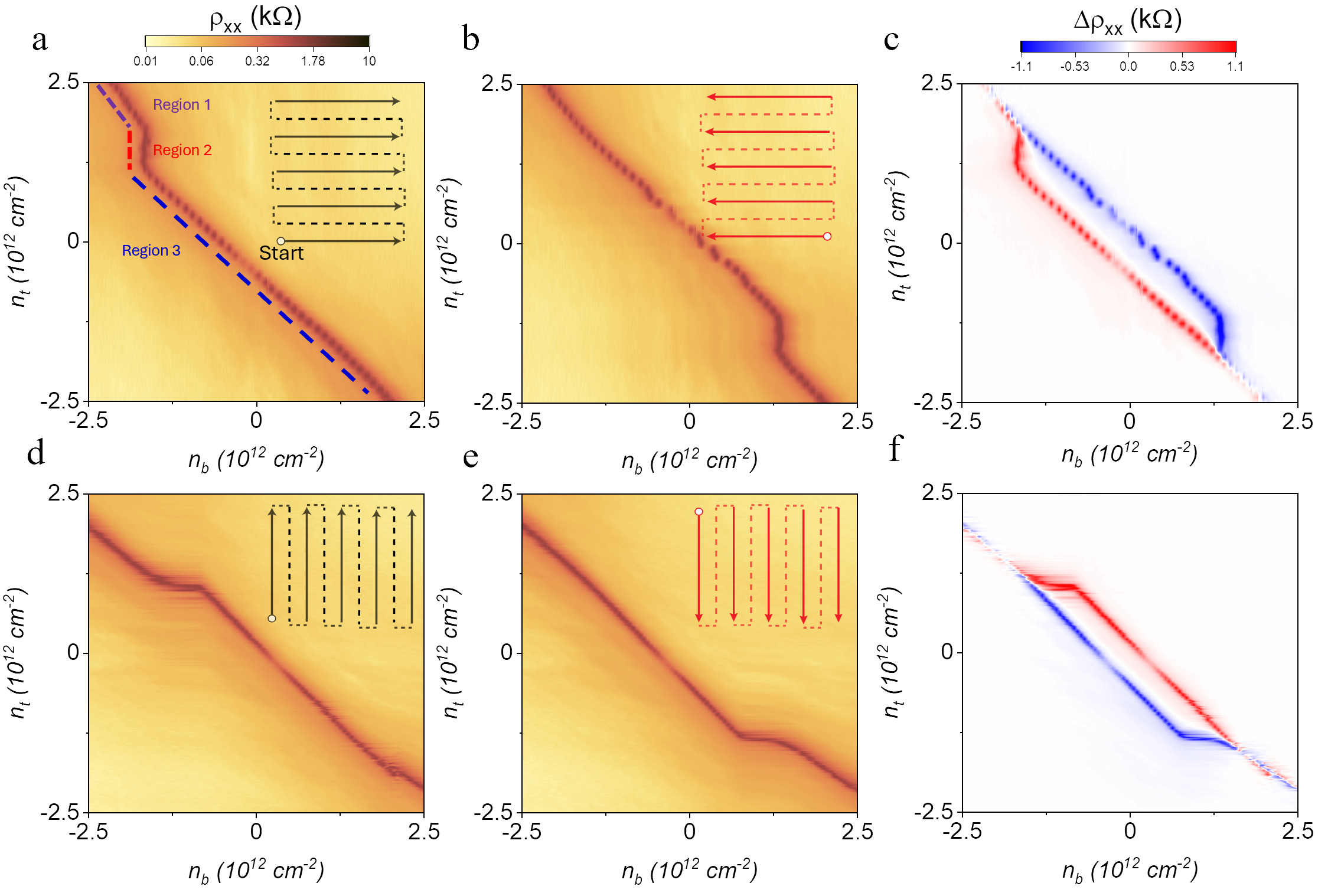}
  \caption{\textbf{Transport characteristics of hBN-encapsulated tetralayer ABCB graphene.}
  \textbf{a,b,} $\rho_{xx}$ as a function of  $n_b$ (fast axis) and $n_t$ (slow axis) for forward (a) and backward (b) sweep directions, measured at $T=5$ K. Insets: the hollow circle marks the sweep starting point; solid lines denote the sweep direction used for plotting data in the respective map. Both forward and backward sweeps were performed at the same rate. 
  \textbf{c,} Difference map of $\rho_{xx}$ between forward and backward sweeps in a and b, highlighting the hysteresis.  
  \textbf{d,e,} $\rho_{xx}$ as a function of $n_t$ (fast axis) and $n_b$ (slow axis) for forward (d) and backward (e) sweeps, acquired by stepping the back gate and sweeping the top gate. 
  \textbf{f,} Difference map of $\rho_{xx}$ between d and e. The hysteresis appears as a characteristic parallelogram-shaped loop, reflecting layer-polarized switching behavior.}
  \label{fig2}
\end{figure}

To further elucidate the hysteresis behavior, we performed two-dimensional gate mapping of $\rho_{xx}$ as a function of $n_t$ and $n_b$, with both forward (up) and backward (down) sweeps acquired at identical sweep rates. Specifically, we swept $n_b$ as the fast axis and stepped $n_t$ as the slow axis, recording $\rho_{xx}(n_b, n_t)$ for both sweep directions (\textbf{Fig.\ref{fig2}a,b}). These maps reveal a single CNP with no satellite features, confirming that neither the top nor bottom hBN is aligned with graphene, consistent with prior observations in misaligned devices \cite{mullan2023mixing, ponomarenko2013cloning}. 

Interestingly, the CNP contour in the up-sweep map (\textbf{Fig.\ref{fig2}a}) divides into three distinct regions. Regions 1 and 2 follow two parallel lines with identical slopes, indicating that both gates contribute proportionately to the carrier modulation in these regimes. Region 2, separating them between $n_t = 1.2 \cdot 10^{12}$ and $1.8 \cdot 10^{12} \mathrm{cm}^{-2}$, shows no back gate dependence, reflecting the onset of layer-specific anomalous screening (LSAS)  \cite{zheng2020unconventional, chen2024anomalous, niu2022giant, niu2025ferroelectricity}. A mirror-symmetric LSAS response is observed in the down-sweep map (\textbf{Fig.\ref{fig2}b}) -- in the negative $n_t$ range ($-1.2$ to $-1.8 \cdot 10^{12} \mathrm{cm}^{-2}$). The difference between the forward and backward sweep maps defines a clear parallelogram-shaped hysteresis loop, \textbf{Fig.\ref{fig2}c}. 

Consistent hysteresis loops are observed across all measured configurations of $\rho_{xx}(n, D)$ (\textbf{Extended Data Fig.\ref{fig_SIUnifromdevice}}), where $n = n_t + n_b$ is the total carrier density and $D = (C_t V_t - C_b V_b)/2\varepsilon_0$ is the displacement field, with $C_{t,b}$ and $V_{t,b}$ denoting the areal capacitance and voltage for the top and bottom gates, respectively. Remarkably, this hysteresis remains robust up to room temperature (\textbf{Extended Data Fig.\ref{fig_Tempraturedependece}}), underscoring its intrinsic nature. Similar behavior is found when sweeping $n_t$ and stepping in $n_b$ (\textbf{Fig.\ref{fig2}d-f}), where the hysteresis features appear with reversed peak positions -- further supporting the layer-specific character of the polarization. We also verified that the CNP peak position is independent of the direction of the stepping axis, whether from electron to hole doping or vice versa.

To systematically explore the behavior of LSAS, we studied how the CNP position depends on the sweep range of $n_t$ and $n_b$. Initially, we performed sweeps of $n_b$ at a fixed $n_t=0$, gradually increasing the sweep range from 0 to $-1.1 \cdot 10^{12} \mathrm{cm}^{-2}$ for each step. The results, shown in \textbf{ Extended Data Fig.\ref{fig_nbRatchet}(a,b)}, reveal that during the forward sweep, the CNP shifts progressively until saturation around $n_t=-0.5 \cdot 10^{12} \mathrm{cm}^{-2}$, whereas during the backward sweep, the CNP position remains largely fixed near $n_t=0$. To further investigate this unconventional hysteresis behavior, we recorded $n_t, n_b$ maps on the electron (\textbf{Extended Data Fig.\ref{fig_nbRatchet}(c,d)}) and hole (\textbf{Extended Data Fig.\ref{fig_nbRatchet}(f,g)}) sides separately. By analyzing the differences between the upward and downward sweeps in these configurations, we observed hysteresis loops whose sizes were defined by $n_b$ sweep ranges, see \textbf{Extended Data Fig.\ref{fig_nbRatchet}(e,h)} for electron- and hole-doped regions, respectively. Similar behavior was observed when sweeping $n_t$ and stepping $n_b$.

\subsection{Charge transfer and microscopic picture of the mechanism}
\begin{figure}[h]
  \centering
  \includegraphics[scale=0.85]{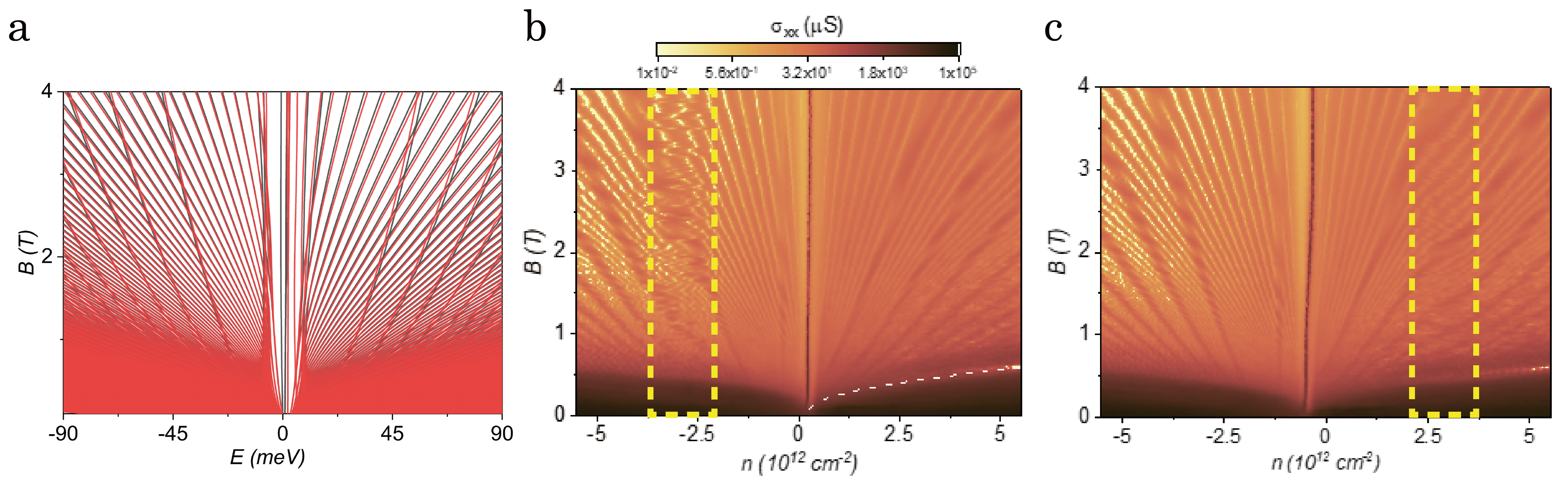}
  \caption{\textbf{Landau level and Hall states in ABCB tetralayer device.} %
  \textbf{a,} Calculated Landau level spectrum of the ABCB tetralayer graphene, the red and black lines correspond to Landau levels from $K^+$ and $K^-$ valleys.  
  \textbf{b,c,} Measured conductivity $\sigma_{xx}$ as a function $n = n_t + n_b$ and $B$ at $D = 0$, $T=5 K$ for forward (b) and backward (c) sweep directions. The yellow dashed boxes indicate the region where structural transition occurs. The white dashed line traces the first transverse magnetic focusing line in our device, see \textbf{Methods, ‘Transport Measurements'} for details.   
}
\label{fig_landaufan}
\end{figure}

Ferroelectric behavior observed in graphene-based heterostructures can sometimes originate from interfacial effects such as electrically switchable dipoles in AB-stacked hBN  \cite{vizner2021interfacial, yasuda2021stacking}, or due to contamination-induced defects within hBN or impurities at the interface between hBN and graphene \cite{cheng2011toward}. Typically, such contamination significantly reduces carrier mobility and suppresses the quantum Hall effect. Our ABCB tetralayer device exhibits high carrier mobility, evidenced by clear observation of Shubnikov-de Haas oscillations, robust quantum Hall effect at relatively low magnetic fields, and magnetic transversal focusing, indicating minimal interface contamination, \textbf{Fig.\ref{fig_landaufan}}. 

To rule out the possibility that hysteresis originates from either the hBN layers or contamination at their interfaces, we fabricated a control device with identical hBN encapsulation but using tetralayer graphene in an ABCA stacking configuration (same graphene flake, but different stacking domain). Measurements of this ABCA-stacked device exhibit conventional transport characteristics without detectable hysteresis between upward and downward gate sweeps in the corresponding   $\rho_{xx}(n_t,n_b)$ maps (\textbf{Extended Data Fig.\ref{figSI_otherdevice}}), confirming that the hysteretic behavior observed in our ABCB devices is intrinsic to the graphene stacking order.

Transport measurements of the ABCB device in the presence of perpendicular magnetic field \textit{B} show that hysteresis in CNP survives at higher B (at least up to $B = 18T$ see \textbf{Extended Data Fig.\ref{fig_SIHighfield}}). Low-field maps conductivity maps $\sigma_{xx}(n, B)$ for the up and the down sweep at $D=0$ are shown \textbf{Fig.\ref{fig_landaufan}b,c}, respectively. The measured Landau fan matches with the calculated Landau fan for the ABCB tetralayer graphene using the SWMC model, further confirming the stacking order (\textbf{Fig.\ref{fig_landaufan}a}, see also \textbf{Methods, `SWMC Model for ABCB graphene system'}). 

The measured $\sigma_{xx}(n, B)$ maps exhibit distinct "fuzzy" regions, highlighted by yellow dashed rectangles in \textbf{Fig.\ref{fig_landaufan}b,c}, indicating areas with bistable electronic behavior. Recent theoretical work by Yang et al. \cite{yang2023atypical} suggests that intrinsic ferroelectricity in multilayer graphene can emerge purely due to stacking-order-induced symmetry breaking, independent of moiré superlattices. Multilayer graphene structures with more than three layers can host polar states, electrically switchable via interlayer sliding. Our experimental observation of robust ferroelectric switching in ABCB tetralayer graphene directly validates this prediction. Here, the polarization reversal arises from gate-induced sliding transitions between distinct stacking configurations, demonstrating that stacking-order control alone, rather than moiré potentials, provides an intrinsic mechanism for ferroelectricity in van der Waals materials. 

\begin{figure}[ht]
  \centering
  \includegraphics[scale=0.8]{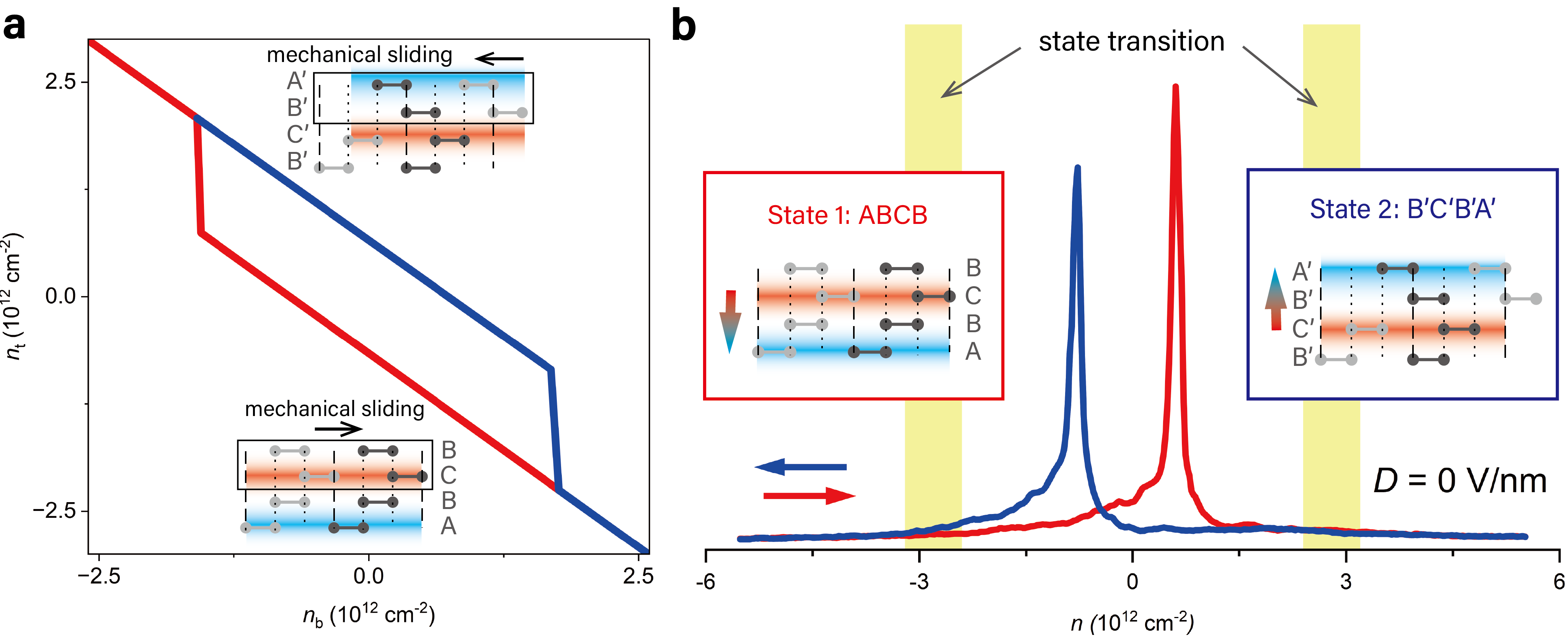}
  \caption{\textbf{Mechanism of hysteresis in ABCB tetralayer graphene.} 
  \textbf{a}, Hysteresis of the charge neutrality point (CNP) as a function of \textit{n}$_t$ and \textit{n}$_b$, caused by a stacking order transition. The red and blue lines trace the graphene resistance during forward and backward sweeps, respectively. Color red and blue denote state 1 and 2 respectively with opposite stacking order. The hysteresis arises from the reversal of charge polarization field in ABCB tetralayer due to a structural transition. As illustrated in \textbf{Fig. \ref{fig1}c}, electrons and holes are localized on C and A layers respectively, creating a net charge polarization within the ABCB graphene. During the gate sweep, the ABCB stacking can transit to B'C'B'A' configuration through mechanical sliding of the two uppermost layers, CB (shown in inset). This effectively reverses both the stacking order and the charge polarization field. 
  \textbf{b}, State transition during gate sweep along the gray dashed line in \textbf{a} at zero displacement field \textit{D} = 0. During gate sweep, the tetralayer graphene transitions between two states with opposite polarization fields: the ABCB stacking order (state 1) and the B'C'B'A' stacking order (state 2). The polarization field (denoted by the blue-red arrow) then contributes oppositely to the CNP shift in each state, leading to characteristic hysteresis. }
  \label{fig3_newmecahnism}
\end{figure}

\textbf{Figure \ref{fig3_newmecahnism}} summarizes the microscopic mechanism underpinning the intrinsic ferroelectric hysteresis observed in ABCB-stacked tetralayer graphene. Panel \textbf{(a)} maps the hysteretic shift in the CNP as a function of top and bottom gate-induced carrier densities. Red and blue lines correspond to the forward and backward backgate gate sweeps, respectively, and each line representing one of two polarization states with opposite stacking orders. The transitions between these two states occur through interlayer mechanical sliding, involving lateral displacement of the two upper graphene layers relative to the lower ones. This stacking transition switches the ABCB configuration to its inverted counterpart, B'C'B'A', thereby reversing the intrinsic polarization field direction, as depicted schematically in the insets. 

Panel \textbf{(b)} illustrates representative measurements of $\rho_{xx} (n)$ at $D = 0$. The pronounced hysteretic shift in the CNP during the forward (red arrow) and backward (blue arrow) gate sweeps highlights the polarization reversal between the ABCB (state 1) and B'C'B'A' (state 2) stacking orders. These two stacking configurations are energetically equivalent but exhibit opposite polarization fields, manifested experimentally as hysteresis. Using the method described in ref.\cite{zheng2020unconventional, niu2022giant}, we estimated the electric polarization in our device $P_{2D}=0.04 \mu C cm^{-2}$. 

 The observed hysteretic behavior and associated ratchet effect suggest that stacking dislocations influence the ferroelectric response in ABCB tetralayer graphene. Systematic gate-induced training experiments (\textbf{Extended Data Fig.\ref{figSI_training}}) demonstrate that repetitive cycling of gate voltages leads to a pronounced shift and narrowing of hysteresis loops, consistent with structural rearrangements and progressive stabilization of local stacking configurations. Additionally, comparative measurements from two nearby regions within the same Hall bar device (\textbf{Extended Data Fig.\ref{fig_ratchet_linetrace}}) exhibit noticeable differences in hysteresis, indicative of spatial variations in local stacking domains dynamically induced by gating. Importantly, these local variations diminish when measurements span the full gate range, suggesting that sufficiently large doping (above a critical value around $2.5-3 \cdot 10^{12} \mathrm{cm}^{-2}$) stabilizes the stacking order uniformly across the device. These observations underscore the complex interplay between stacking-order dynamics, domain formation, and intrinsic ferroelectricity in ABCB graphene structures. 

Our experimental results demonstrate intrinsic ferroelectricity in ABCB-stacked tetralayer graphene arising directly from stacking-order-induced symmetry breaking. Through systematic gate-dependent transport measurements, we revealed robust hysteresis and a pronounced ratchet effect associated with reversible interlayer sliding and spontaneous polarization switching. Spatial variations and gate-induced domain formation further illustrate the complexity and tunability of stacking-order domains in multilayer graphene systems. This highlights stacking-order control as an intrinsic and versatile mechanism for engineering ferroelectricity in elemental van der Waals materials. These findings not only confirm recent theoretical predictions but also open promising avenues for exploring electrically switchable polarization phenomena and functional device architectures based on graphene multilayers.




\section{Methods}\label{sec2}
\subsection{Device Fabrication}
Our hBN/tetralayer graphene/hBN heterostructures were stacked using the dry transfer method \cite{mishchenko2014twist, wang2013one}. The tetralayer graphene and hBN flakes were first exfoliated in the 290 nm $Si/SiO_2$ wafer, then the hBN/ABCB flakes were picked sequentially by the polymer of polydimethylsiloxane and polymethylmethacrylate (PMMA) and then the vdW heterostructure was dropped into the bottom hBN flake.  To prevent the crystallographic alignment of the hBN and ABCB flakes and make band reconstruction energies inaccessible in transport measurements, we intentionally misaligned both the top and bottom hBN flakes. Electrical contacts were exposed by electron-beam (e-beam) lithography in PMMA resist, the top hBN layers were etched using reactive ion etching, and the metal contacts Cr/Au (3 nm/80 nm) were deposited using the e-beam evaporator. To deposit the top gate, another round of e-beam lithography, deposition, and plasma etching was used to define the Hall bar geometry.

\subsection{Stacking order of tetralayer graphene}
Graphene with four layers typically exhibits an ABAB stacking sequence, while ABCA and ABCB configurations are less common and possess lower stability. To unambiguously identify the stacking order, we performed atomic force microscopy (AFM) and Raman spectroscopy. AFM raster scans were acquired using a Bruker Dimension Icon with Scanasyst-Fluid+ probes with a silicon tip. The Raman maps were recorded using a Horiba Xplora+ confocal Raman microscope using an excitation wavelength of 532 nm focused with a 100x objective, a 1800 gr/mm grating with a spatial resolution of 500 nm, and an integration time of 10 s per pixel. The laser power was set to 2 mW. The results are summarized in  \textbf{Extended data Fig.\ref{figSI_AFM_Raman}}. 

First, we measured the thickness of the graphene flake as exfoliated on the Si/SiO$_2$ substrate before encapsulation extracting a height profile perpendicular to the edge of the graphene flake. We fitted the profile with a step-like function, yielding a thickness of 1.38 $\pm$ 0.04 nm (\textbf{Extended data Fig.\ref{figSI_AFM_Raman}b}), which, considering a thickness of 0.34 nm for monolayer graphene, perfectly matches the thickness of a 4L graphene. In addition, we performed hyperspectral Raman spectroscopy mapping on the encapsulated flake before the fabrication of the hall bars. We fitted each Raman spectrum using a series of Lorentzian functions, using 1 Lorentzian for the G band and 3 Lorentzians to properly characterize the 2D band. In \textbf{Extended data Fig.\ref{figSI_AFM_Raman}c} we show the intensity, position, and full-width half-maximum (FWHM) hyperspectral maps resulting from the fits, where the ABCA and ABCB regions are clearly identified. To complement the maps, we include selected spectra from each region in \textbf{Extended data Fig.\ref{figSI_AFM_Raman}d-f}. The position of the G peak depends on the stacking order: ABCA sequence is shifted to lower wavenumbers compared to ABCB and ABAB \cite{wirth2022experimental, beitner2023mid, zhou2024inversion}. In our case, we find two regions centered at $\sim$ 1577 cm$^{-1}$ and $\sim$ 1580 cm$^{-1}$, which we attribute to ABCA and ABCB stacking order. Moreover, the shape of the 2D band ultimately allows to unequivocally identify the stacking order: ABAB sequence shows a symmetric shape, ABCA presents a clear asymmetric shape, and ABCB exhibits a weaker asymmetry but it is still present \cite{wirth2022experimental, beitner2023mid, zhou2024inversion}. In our measurements, we observe asymmetric 2D bands, with strong and weak asymmetry, that we can assign to ABCA and ABCB stacking order respectively. These results are in perfect agreement with previously reported literature \cite{wirth2022experimental, beitner2023mid, zhou2024inversion}.

\subsection{Transport Measurement}
The longitudinal and Hall voltages were recorded using an SR830 lock-in amplifier by applying a small a.c. currents of about 10-100 nA at a frequency of 33.3 Hz in a commercial cryostat equipped with a superconducting magnet. Gate voltages were applied using the DC voltage source meter (Keithley 2614B). 

The magnetic field ($B_f$) required for the focusing of electrons at distance $L$ in  the multilayer graphene \cite{tsoi1974focusing,tsoi1999studying,taychatanapat2013electrically}  can be written as:
\begin{equation}
    B_f^{(p)}= \left( \frac{2 \kappa_f \hbar}{eL} \right) p
\end{equation}
where $p-1$ is the number of reflections before the electron reaches from injector the collector voltage probe, $\kappa_f = \sqrt{4\pi n/g_vg_s}$ is the Fermi momentum with $g_s=g_v=2$ being the spin and valley degeneracy in the graphene system and n being the charge carrier density and $\hbar$ corresponds to the reduced Planck's constant.

\subsection{DFT calculations}
The electronic band structure and projected bands on each graphene layer were calculated using first-principles calculations performed with the Vienna ab initio simulation package using the projector-augmented wave method \cite{kresse1999ultrasoft}. The generalized gradient approximation parametrized by Perdew–Burke–Ernzerhof was used for exchange-correlation functionals \cite{perdew1996generalized}.  Brillouin-zone integration was performed using a $16\times16\times1$ M-centered k-point mesh. The kinetic energy cut-off point for the plane wave basis set was $500eV$.

\subsection{SWMC Model for ABCB graphene system}
To compute the electronic band structure of ABCB stacked graphene, we adapted the Slonczewski-Weiss-McClure (SWMcC) parametrization \cite{slonczewski1958band, mcclure1957band, sarsfield2024substratetemperaturemagneticfield} combined with the self-consistent screening potential induced by the top and bottom gates charge carriers $n_t$ and $n_b$. The hopping amplitudes we used are $\gamma_0=3.16eV$, $\gamma_1=390meV$, $\gamma_2=-17meV$, $\gamma_3=315meV$, $\gamma_4=70meV$, $\gamma_5=30meV$ and $\Delta=-5meV$ were adapted from the earlier work \cite{yin2019dimensional}.

\bibliography{sn-bibliography}


\backmatter

\section*{Data Availability}
Relevant data are available from the corresponding authors on request.

\section*{Acknowledgments}
This study was supported by the European Research Council (ERC) under the European Union’s Horizon 2020 research and innovation program (Grant Agreement No. 865590) and the Research Council UK [BB/X003736/1].  A.S. would like to acknowledge the support from Dean's doctoral scholarship provided by the University of Manchester. S.X. acknowledges support from the National Natural Science Foundation of China (Grant No. 12274354). Q.Y. acknowledges the funding from Royal Society University Research Fellowship URF\textbackslash R1\textbackslash221096 and UK Research and Innovation Grant [EP/X017575/1].

\section*{Author Contributions}
A.M. planned the project and conceived the idea and experiments. A.S. performed the transport measurements and data analysis under the supervision of A.M. S.X. fabricated the device. A.S., P.J.S., and S.S. performed the tight-binding and DFT calculations under the supervision of A.M. and V.I.F. P.D.N., N.K., and Z.W. did the optical spectroscopy measurements and characterizations. A.S. and A.M. wrote the manuscript with inputs from P.J.S. and V.F. All authors discussed the results and commented on the paper.

\section*{Competing Interests}
The authors declare no competing interests.

\newpage

\begin{appendices}

\section{Extended Data}\label{secA1}
\renewcommand{\figurename}{Extended Data Fig.}


\begin{figure*}[ht]
  \centering
  \includegraphics[width=\columnwidth]{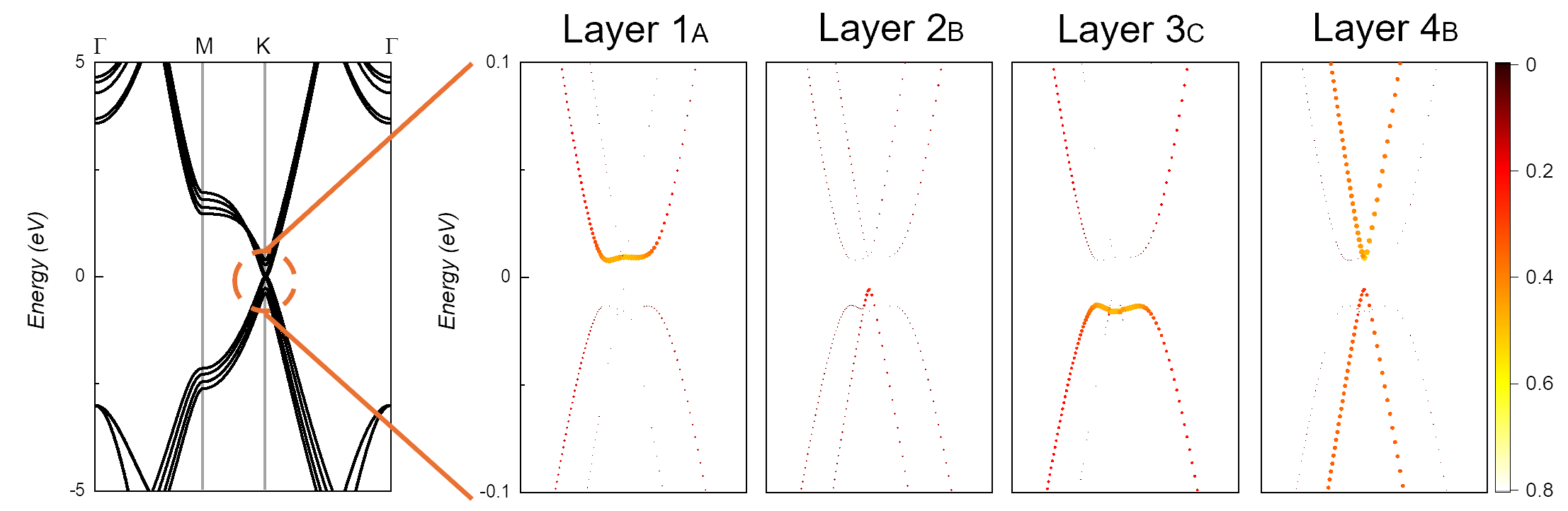}
  \caption{\textbf{DFT calculated layer projected bandstructure of ABCB stacked tetralayer graphene band structure.} Calculated energy dispersion of tetralayer ABCB stacked graphene as a function of in-plane momentum along the high symmetry line ($\Gamma - M - K - \Gamma$) and energy (vertical axis). The enlarged view shows the density of states projected in layers near the Fermi surface of the ABCB graphene system. Yellow color indicates a high probability of the density of the states in different energy bands. The Fermi level is set to be $0eV$.  }
  \label{figSI_dispersion}
\end{figure*}

\begin{figure*}[ht]
  \centering
  \includegraphics[width=\columnwidth]{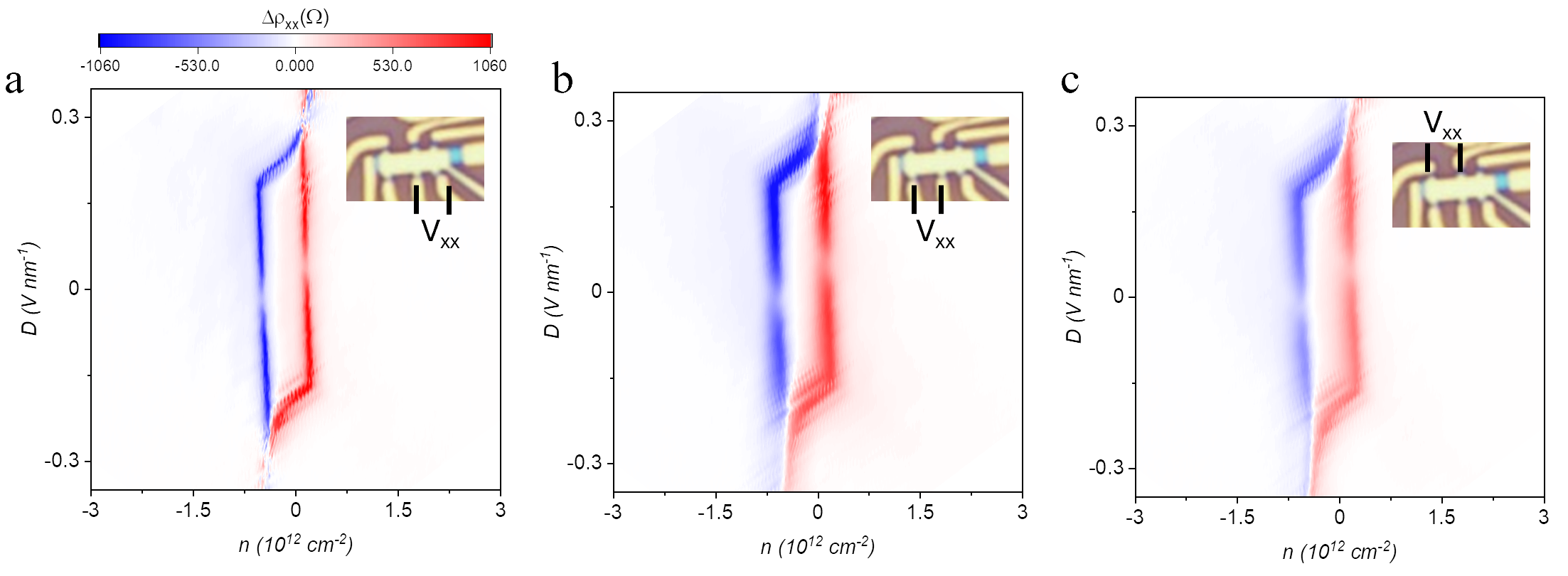}
  \caption{\textbf{Uniform hysteretic behavior across all measured configurations in the device. a-c,} The difference of resistivity maps $\rho_{xx}(n, D)$ for up and down sweep map measured by sweeping the $n_t$ and stepping in $n_b$ for all the measured configurations. Insets: optical micrographs of the device highlighting the contact pairs with which the measurements were taken.}
  \label{fig_SIUnifromdevice}
\end{figure*}

\begin{figure*}[ht]
  \centering
  \includegraphics[width=\columnwidth]{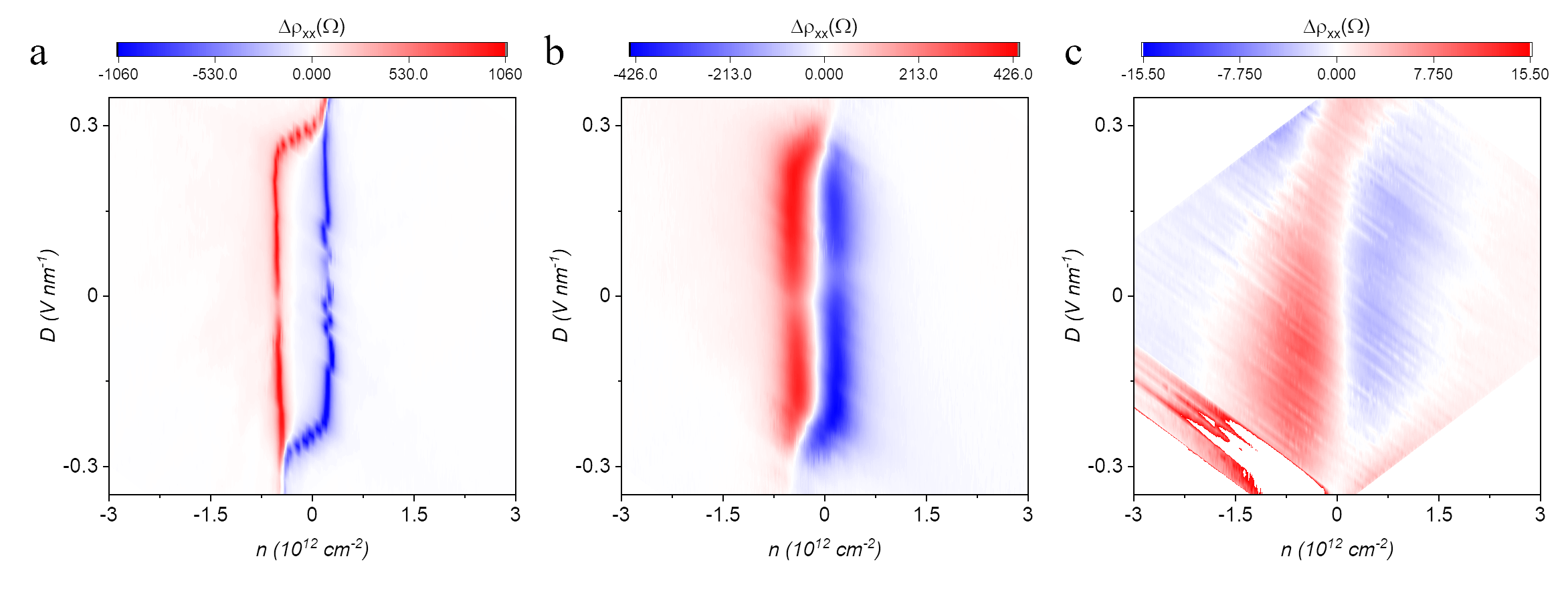}
  \caption{\textbf{Temperature dependence of the hysteric behavior.} The difference of resistivity maps $\rho_{xx}(n, D)$ for forward and backward sweep map measured by sweeping the $n_b$ and stepping in $n_t$ at different temperatures of \textbf{a,} T=5K \textbf{b,} T=50K  and \textbf{c,} T=300K.}
  \label{fig_Tempraturedependece}
\end{figure*}

\begin{figure*}[ht]
  \centering
  \includegraphics[width=\columnwidth]{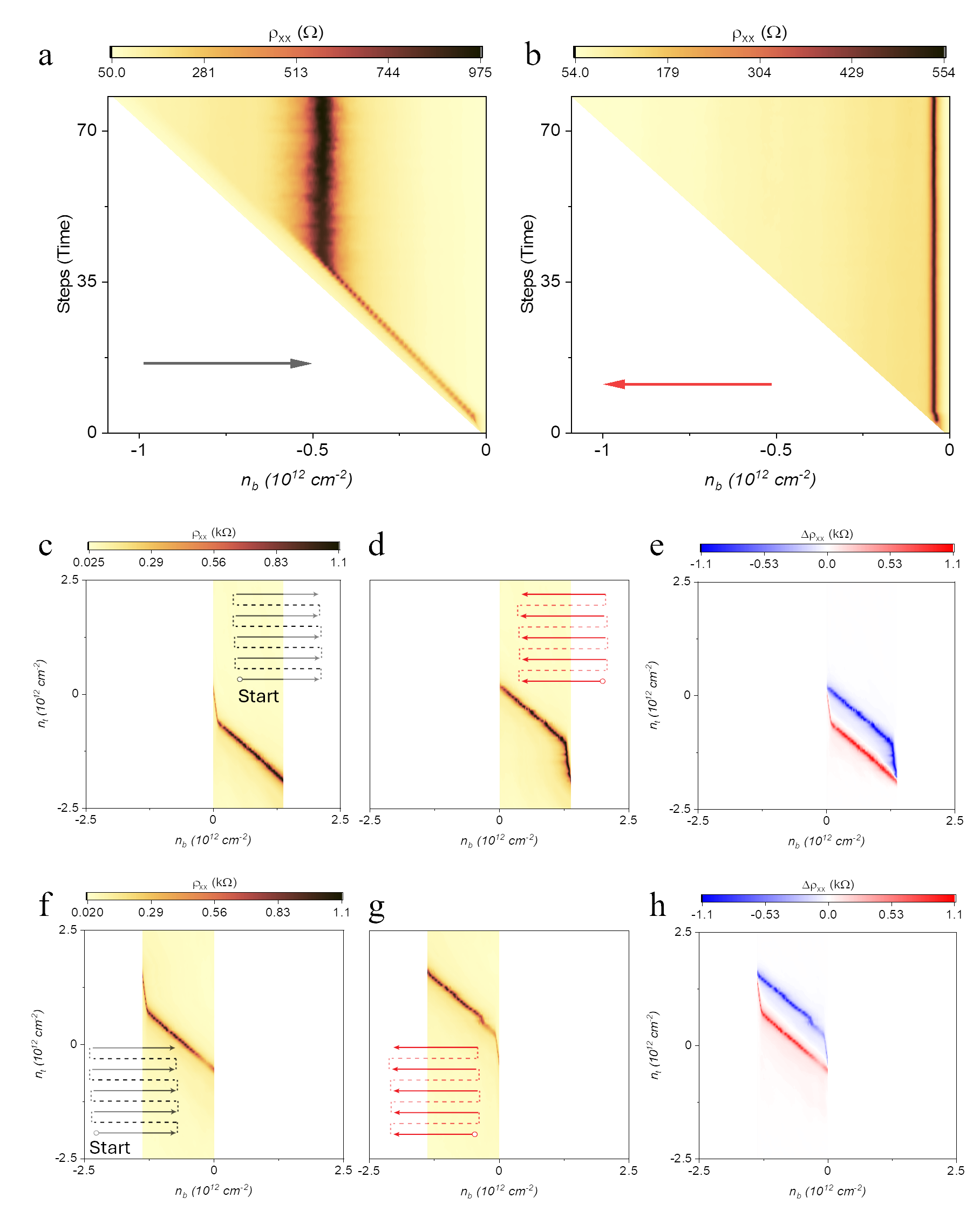}
  \caption{\textbf{Ratchet effect(shifting hysteresis) in $n_b$ sweeps and $n_t$ steps. a-b,} $\rho_{xx}$ as a function of $n_b$ at $n_t=0$ for forward and backward sweep, with directions indicated by the black and red arrows. The data was taken by changing the sweep range from 0 to $-1.1\times 10^{12}cm^{-2}$ at each step. \textbf{c-d,} Map of $\rho_{xx}$ ($n_b, n_t$) for forward and backward sweep of $n_b$ from $0$ to $1.38\times10^{12}cm^{-2}$ with $n_t$ stepped from $-2.62\times10^{12}cm^{-2}$ to $2.62\times10^{12}cm^{-2}$. \textbf{e,} The map of difference resistivity between \textbf{c,} forward sweep and \textbf{d,} backward sweep. \textbf{f-h,} Similar maps in \textbf{c-e,} obtained by changing the sweep range from $-1.38\times10^{12}cm^{-2}$ to $0\times10^{12}cm^{-2}$ with $n_t$ stepped from $-2.62\times10^{12}cm^{-2}$ to $2.62\times10^{12}cm^{-2}$. }
  \label{fig_nbRatchet}
\end{figure*}

\begin{figure*}[ht]
  \centering
  \includegraphics[width=\columnwidth]{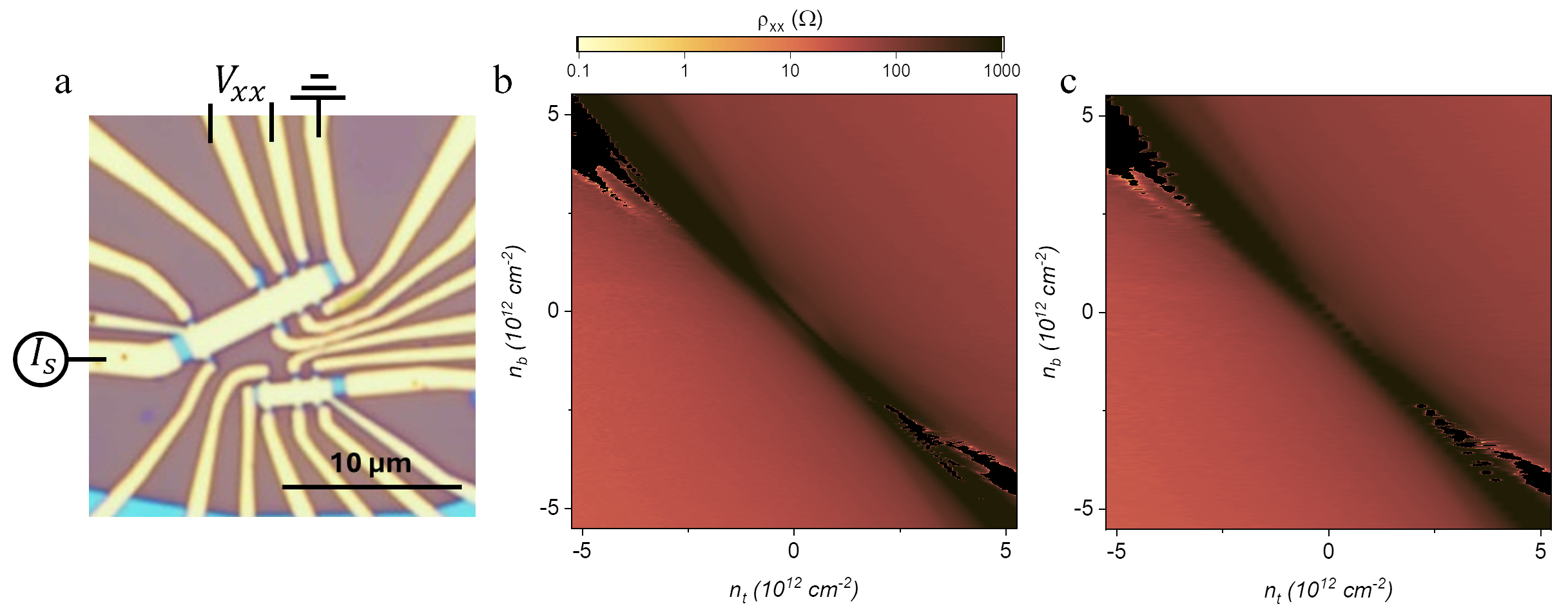}
  \caption{\textbf{Transport properties of another device fabricated from the same hBN stack. a,} Optical micrograph of the device. The annotations show the measurement setup used in the measurements of big hall bar. $I_s$ is the a.c. source, $V_{xx}$ is the potential drop measured across the longitudinal xx pair. The scale bar is $10\mu m$. \textbf{b-c,} $\rho_{xx}$ as a function of  $n_b$ and $n_t$ for (\textbf{b}) forward and (\textbf{c}) backward sweep (the backward sweep was measured at twice the sweep rate of forward speed) measured at T=5K, the maps were measured by sweeping $n_t$ (fast axis) and stepping in $n_b$ (slow axis). }
  \label{figSI_otherdevice}
\end{figure*}

\begin{figure*}[ht]
  \centering
  \includegraphics[width=\columnwidth]{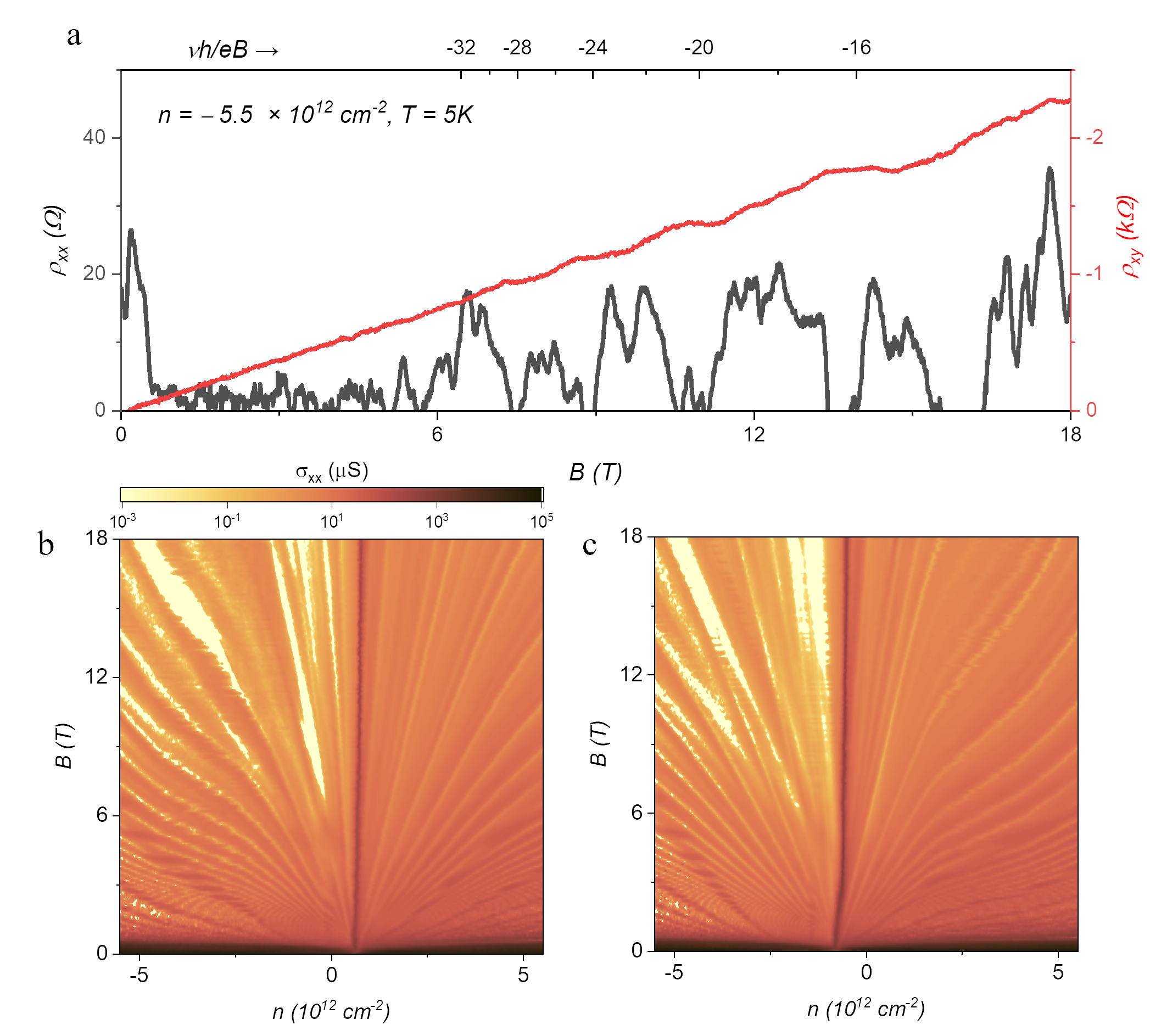}
  \caption{\textbf{Quantum hall effect and Landau fan in ABCB graphene at higher magnetic fields. a,} $\rho_{xx}$ and $\rho_{xy}$ (transversal resistivity) as a function of B, measured at $5K$, $D=0$, and $n=-5.5\times10^{12}cm^{-2}$. The top axis shows the filling factor $\nu$. Conductivity ($\sigma_{xx}$) maps as a function of $n$ and $B$ (up to $18T$) at $T=5K$ for \textbf{b,} forward and \textbf{c,} backward sweep. }
  \label{fig_SIHighfield}
\end{figure*}

\begin{figure*}[ht]
  \centering
  \includegraphics[width=0.8\columnwidth]{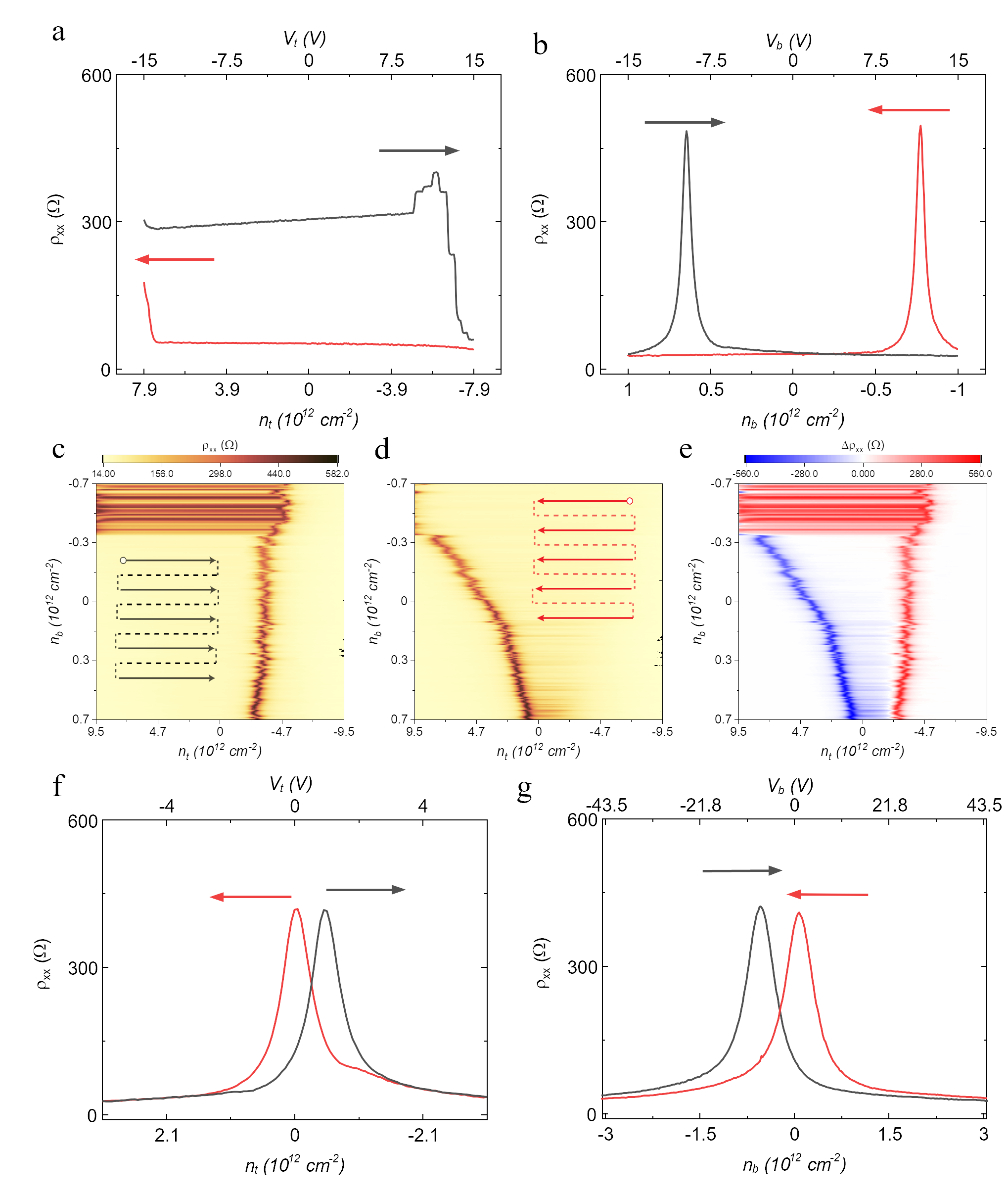}
  \caption{\textbf{Training the $n_b$ and $n_t$ at T=50K. a-b,}  $\rho_{xx}$ as a function of $n_t$ measured at  $n_b=0$ and $n_b$ measured at and $n_t=0$ before the training, the top axis shows the applied top-back gates voltages. The black and red arrow indicate the forward and backward gate sweep directions. \textbf{c-d},$\rho_{xx}$ as a function of  $n_t$ and $n_b$ for forward and backward sweep. Insets: sweeping directions where the start point for the measurement is denoted by the hollow circle. The maps are taken by stepping in the back gate bias (slow axis) and sweeping the top gate bias (fast axis), both forward and backward sweeps were swept at the same sweep rate. \textbf{e}, The difference in the resistivity measured in \textbf{c} and \textbf{d}. \textbf{e-f}, $\rho_{xx}$ as a function of $n_t$ measured at  $n_b=0$ (e), and $n_b$ measured at $n_t=0$ (f) after the training. }
  \label{figSI_training}
\end{figure*}

\begin{figure*}[ht]
  \centering
  \includegraphics[width=\columnwidth]{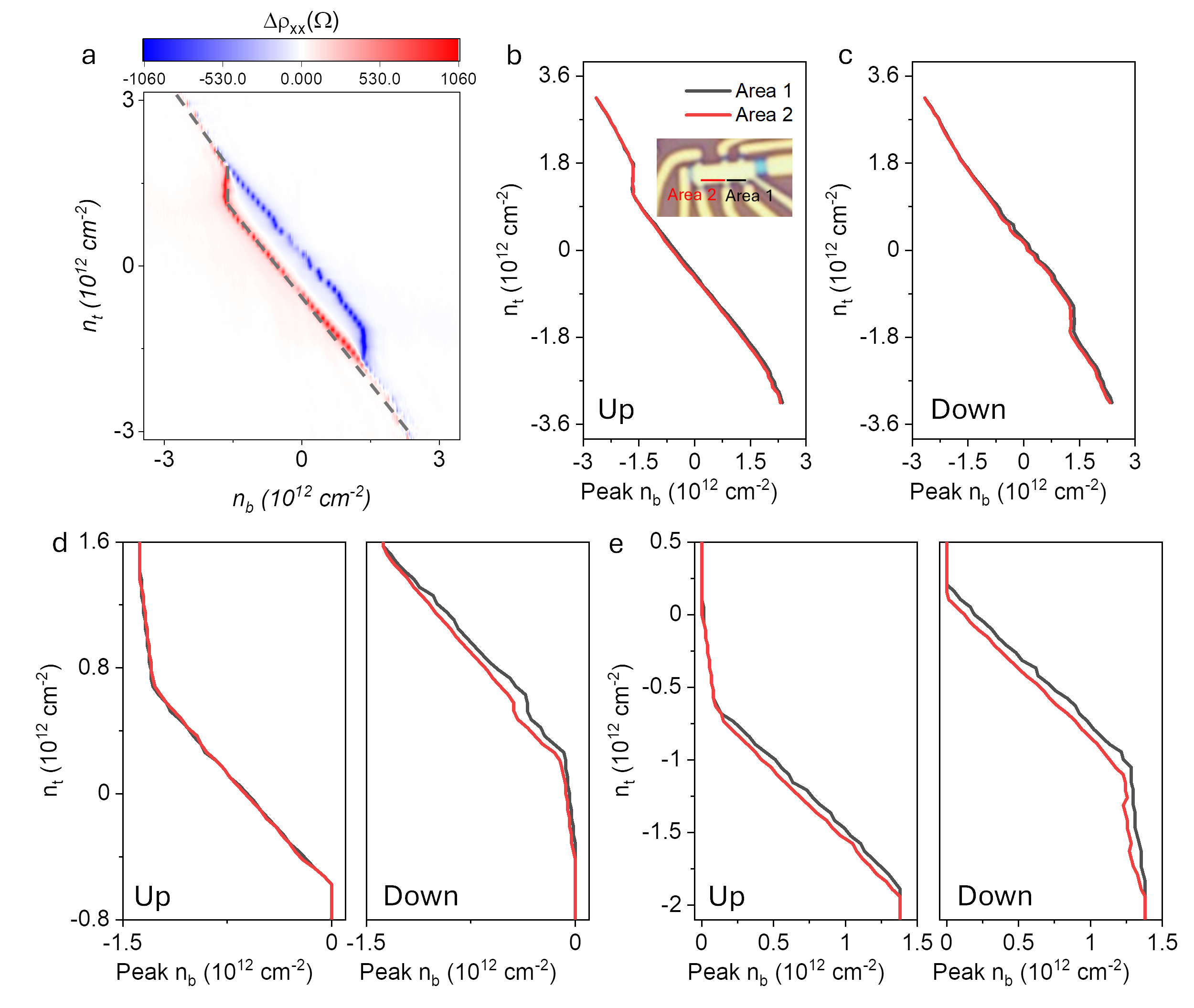}
  \caption{\textbf{Ratchet effect in different areas of the device. a,} The difference map between the up and down sweep of $\rho_{xx}(n_b,n_t)$ maps is measured by sweeping the $n_b$ and stepping in $n_t$ (the figure is same as in main text \textbf{Fig.\ref{fig2}c}). The dashed line indicates the CNP peak position.  \textbf{b-c,} The line trace of the peak position as a function of $n_t$ steps is shown for two distinct areas within the device during both the upward and downward sweeps. Inset: an optical micrograph of the device highlighting the measurement areas. Line traces of CNP for upward and downward sweeps in various areas within the device were conducted over $n_b$ sweep ranges from \textbf{(d)} $0$ to $1.38\times10^{12}cm^{-2}$ and \textbf{(e)} $-1.38\times10^{12}cm^{-2}$ to $0$, with $n_t$ stepping from $-2.62\times10^{12}cm^{-2}$ to $2.62\times10^{12}cm^{-2}$. }
  \label{fig_ratchet_linetrace}
\end{figure*}

\begin{figure*}[ht]
  \centering
  \includegraphics[width=0.9\columnwidth]{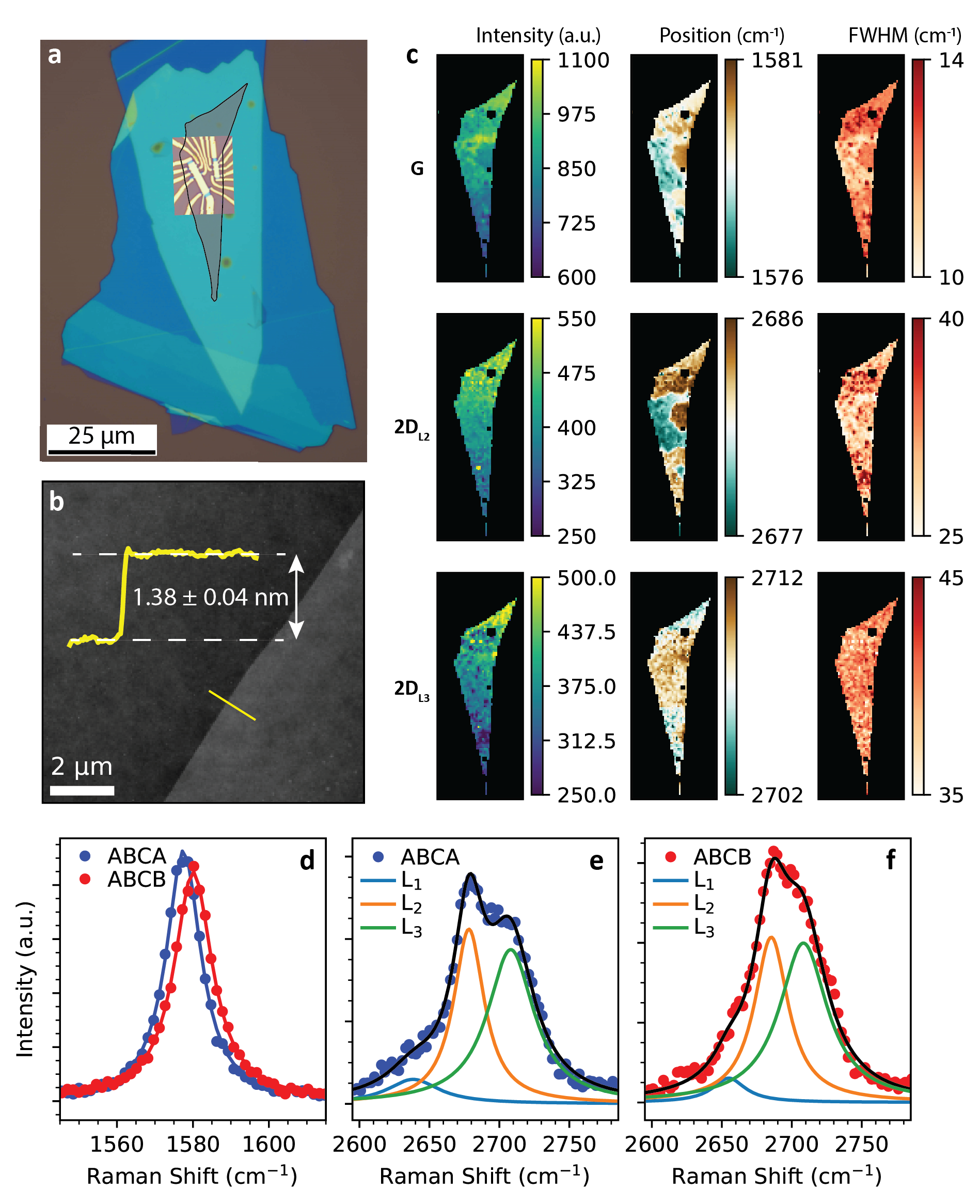}
  \caption{\textbf{Stacking order of tetralayer graphene analyzed by atomic force microscopy and Raman spectroscopy. a,} Optical image of the encapsulated tetralayer graphene. The position of the graphene flake is highlighted with a continuous black line and darker color. The hall-bars are overlapped to indicate their position in the flake. \textbf{b,} Topography of the graphene's edge flake exfoliated on the Si/SiO$_2$ substrate before encapsulation. The inset represents a profile perpendicular to the flake's edge, indicating a thickness of $\sim$ 1.4 nm. \textbf{c,}  Fitting results of the hyperspectral Raman spectroscopy maps for the G peak and the 2D peak (L$_2$ and L$_3$). \textbf{d,} G band fitting for the ABCA and ABCB regions. \textbf{e,f,} 2D band fitting for the ABCA and ABCB regions, respectively, indicating the three components (L$_1$, L$_2$, and L$_3$) used for the fitting. The asymmetry in the bands is clear in both systems and stronger in ABCA. }
  \label{figSI_AFM_Raman}
\end{figure*}


\end{appendices}

\end{document}